\documentstyle[prl,aps,epsf,floats]{revtex}
\input epsf

\begin{document}
\twocolumn[\hsize\textwidth\columnwidth\hsize\csname@twocolumnfalse\endcsname

\draft

\title{Spin Degree of Freedom in a Two-Dimensional Electron Liquid}
\author{Tohru Okamoto,$^1$ Kunio Hosoya,$^1$ Shinji Kawaji,$^1$ and Atsuo Yagi$^2$}
\address{$^1$Department of Physics, Gakushuin University, Mejiro, Toshima-ku, Tokyo 171-8588, Japan}
\address{$^2$NPC Ltd., Shiobara-cho, Tochigi 329-2811, Japan}
\date{October 16, 1998}
\maketitle

\begin{abstract}
We have investigated correlation between spin polarization and magnetotransport in a high mobility silicon inversion layer which shows the metal-insulator transition.
Increase in the resistivity in a parallel magnetic field reaches saturation at the critical field for the full polarization evaluated from an analysis of low-field Shubnikov-de Haas oscillations.
By rotating the sample at various total strength of the magnetic field, we found that the normal component of the magnetic field at minima in the diagonal resistivity increases linearly with the concentration of ``spin-up'' electrons.

\end{abstract}

]

A metal-insulator transition (MIT) observed in Si metal-oxide-semiconductor field-effect transistors \cite{KRAV1,POPO} (Si-MOSFET's) and other systems \cite{SIGE1,SIGE2,GAAS1,GAAS2} attracts a great deal of attention since it seems to contradict an important result of the scaling theory by Abrahams {\it et al.} \cite{GANG4} that the conductance of a disordered two-dimensional (2D) system at zero magnetic field goes to zero for $ T \rightarrow 0 $.
In the metallic phase in Si-MOSFET's with high peak electron mobilities of $ \mu _{ \rm peak} \gtrsim 2~{ \rm m }^{2} / { \rm V~s} $, the diagonal resistivity $ \rho_{ xx } $ shows a sharp drop with decreasing temperature from about $ 2~{\rm K} $ \cite{KRAV1}.
Recent experiments \cite{KRAV2,PUD1} show that magnetic fields applied parallel to the 2D plane suppress the low temperature metallic conduction in Si-MOSFET's.
Since the parallel magnetic field does not couple the orbital motion of electrons, this fact suggests an important role of the spin of electrons.
However, the mechanism of the conduction in the anomalous metallic phase is not clear yet.

The 2D systems that show the MIT \cite{KRAV1,POPO,SIGE1,SIGE2,GAAS1,GAAS2} are characterized by strong Coulomb interaction between electrons.
The mean Coulomb energy per electron $ U = ( \pi N _ {s} )^{ 1 / 2 } e ^ { 2 } / 4 \pi \varepsilon _ { 0 } \kappa $ is larger than the mean kinetic energy $ K = \pi \hbar ^ { 2 } N _ { s } / m ^ { \ast } $ by an order of the magnitude around the critical point for the MIT.
Here, $ N _ {s} $ is the electron concentration, $ \kappa $ is the relative dielectric constant at the interface, and $ m ^ { \ast } $ is the effective mass of electron.
It is estimated that $ U = 120~{ \rm K } $, $ K = 14~{ \rm K } $ and the ratio $ r _ { s } = U / K = 8.3 $ for $ \kappa = 7.7 $ and $ m ^ { \ast } = 0.19 m _ { e } $ at $ N _ {s} = 1 \times 10 ^ { 15 } {\rm m } ^ { -2 } $ in Si-MOSFET's.
The ground state of the insulating phase of high mobility Si-MOSFET's is considered to be a pinned Wigner solid (WS) \cite{PUD2,CHUI}.
Magnetic field dependence of the thermal activation energy observed for various angles of the magnetic field was essentially explained by a model based on magnetic interactions in the pinned WS \cite{OKA1,OKA2}.
Although the quantum fluctuations change the 2D system into a liquid at higher-$ N _ {s} $, electron-electron ({\it e-e}) interaction is expected to be still important.

In the conduction band of silicon, the spin-orbit interaction is negligible and the spin polarization $ p = ( N _ { \uparrow } - N _ { \downarrow } ) / N _ { s } $ can always be given in the direction to the magnetic field.
Here $ N _ { \uparrow } $ and $ N _ { \downarrow } $ are the concentrations of electrons having an up spin and a down spin, respectively $ ( N _ { s } = N _ { \uparrow } + N _ { \downarrow } ) $.
In the present work, we investigate the low temperature conduction in a high mobility Si-MOSFET for various values of $ p $.

The Si-MOSFET sample used has a peak electron mobility of $ \mu _{ \rm peak} = 2.4~{ \rm m }^{2} / { \rm V~s} $ at $ N _ { s } = 4 \times 10 ^ { 15 } { \rm m } ^ { -2 } $ and $ T = 0.3~{ \rm K } $.
It has a Hall-bar geometry of total length 3~mm and width 0.3~mm.
The estimated SiO$_2$ layer thickness is 98~nm.
The diagonal resistivity $ \rho _ { xx } $ as well as the Hall resistivity $ \rho _ { xy } $ was obtained from the linear relationship in the { \it I-V } characteristics using a four-probe dc method.
The source-drain current and the potential difference across the two potential probes separated by 1.5~mm were limited below 10~nA and 0.4~mV, respectively.
The sample was mounted on a rotatory thermal stage in a dilution refrigerator together with a GaAs Hall generator and a RuO$_2$ resistance thermometer calibrated in magnetic fields \cite{OKA2}.

\begin{figure}
\centerline{
\epsfysize=77mm
\epsfbox{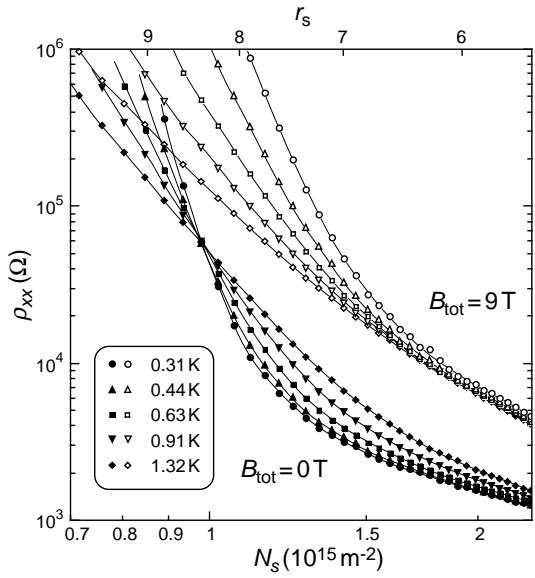}
}
\vspace{4mm}
\caption{ Diagonal resistivity as a function of electron concentration for different temperatures in a zero magnetic field (closed symbols) and in a parallel magnetic field of $ 9~{\rm T} $ (open symbols). }
\label{FGMIT}
\end{figure}

As shown in Fig.~\ref{FGMIT},
the MIT is clearly observed in a zero magnetic field.
The critical value of $ \rho _ { xx } $ at the MIT is $ 55~{ \rm k } \Omega $.
It is close to $ \rho _ { c } \sim 2 h / e ^ { 2 } $ reported by Kravchenko {\it et al.} \cite{KRAV1}.
A high magnetic field of 9~T parallel to the 2D plane drastically increases $ \rho _ { xx } $ and eliminates the positive temperature dependence even in the low-$ \rho _ { xx } $ region \cite{HT}.
The disappearance of the MIT due to the parallel magnetic field seems to resemble that induced by increasing disorder reported by Popovi\'{c} {\it et al.} \cite{POPO}.

In non-interacting degenerate 2D Fermi gases, the spin polarization $ p $ increases linearly with the total strength $ B _ { \rm tot } $ of the magnetic field for $ p < 1 $ and reaches $ p = 1 $ at the critical magnetic field of $ B_{ c } = 2 \pi \hbar ^ { 2 } N _ { s } / \mu _ { B } g _ { v } g ^ { \ast } m ^ { \ast } $.
Here, $ \mu _ { B } $ is the Bohr magneton $ ( = \hbar e / 2 m _ { e } ) $.
The valley degeneracy $ g _ { v } $ is 2 on the (001) surface of silicon and the effective $ g $-factor $ g ^ { \ast } $ is 2.0 in the conduction band in silicon \cite{KODERA}.
According to Landau's Fermi liquid theory, the {\it e-e} interaction changes the effective $ g $-factor and mass of electrons (or quasiparticles) into enhanced values denoted by $ g _ { \rm FL } $ and $ m _ { \rm FL } $, and reduces the critical magnetic field.

\begin{figure}
\centerline{
\epsfysize=58mm
\epsfbox{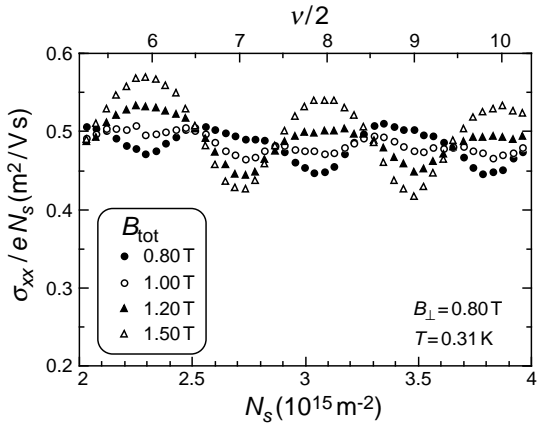}
}
\vspace{4mm}
\caption{ Shubnikov-de Haas oscillations at $ B _ { \perp } = 0.80~{\rm T } $ and $ T = 0.31~{\rm K } $ for various $ B _ { \rm tot } $.
}
\label{FGOS}
\end{figure}

The product of $ g _ { \rm FL } $ and $ m _ { \rm FL } $ in (001) Si-MOSFET's can be determined from the Shubnikov-de Haas oscillations in tilted magnetic fields \cite{FS}.
Figure~\ref{FGOS} shows the diagonal conductivity $ \sigma _ { xx } = \rho _ { xx } / ( \rho _ { xx } {} ^ { 2 } + \rho _ { xy } {} ^ { 2 } ) $ divided by $ e N _ { s } $ for various angles of the magnetic field to the 2D plane.
The normal component $ B _ { \perp } $ of the magnetic field is fixed at $ 0.80~{ \rm T } $.
The upper abscissa is given by a half of the Landau level filling factor $ \nu = N _ { s } \phi _ { 0 } / B _ { \perp } $ ($ \phi _ { 0 } = h / e $) taking into account the valley degeneracy $ g _ { v } = 2 $.
We ignore the valley splitting energy since no indication is observed at $ \nu = odd $.
For $ B _ { \rm tot } = 0.80~{ \rm T } $, $ \sigma _ { xx } / e N _ { s } $ takes minima at even-numbers of $ \nu / 2 $ where the Fermi energy is located at the middle of a gap between Landau levels having different Landau quantum number and spin index.
This energy gap $ \Delta E _ { even } = \hbar e B _ { \perp } / m _ { \rm FL } - g _ { \rm FL } \mu _ { B } B _ { \rm tot } $ decreases with $ B _ { \rm tot } $ and the minima in $ \sigma _ { xx } / e N _ { s } $ at even-numbers of $ \nu / 2 $ fade away at higher-$ B _ { \rm tot } $ due to the level broadening.
On the other hand, the spin splitting $ \Delta E _ { odd } = g _ { \rm FL } \mu _ { B } B _ { \rm tot } $ increases with $ B _ { \rm tot } $ and the minima in $ \sigma _ { xx } / e N _ { s } $ at odd-numbers of $ \nu / 2 $ become clearer.
We have evaluated the critical angle $ \theta _ { c } $ of the magnetic field to the 2D plane where $ \Delta E _ { odd } $ becomes equal to $ \Delta E _ { even } $ from comparing the depth of the minimum in $ \sigma _ { xx } / e N _ { s } $ at an odd-number of $ \nu / 2 $ with the average of those at $ \nu / 2 - 1 $ and $ \nu / 2 + 1 $ for various values of $ B _ { \rm tot } $.
We assumed that the level broadening is a smooth function or independent of $ \nu $.
Thus we have $ g _ { \rm FL } m _ { \rm FL } = m _ { e } B _ { \perp } / B _ { \rm tot } = m _ { e } \sin \theta _ { c } $.

\begin{figure}
\centerline{
\epsfysize=79mm
\epsfbox{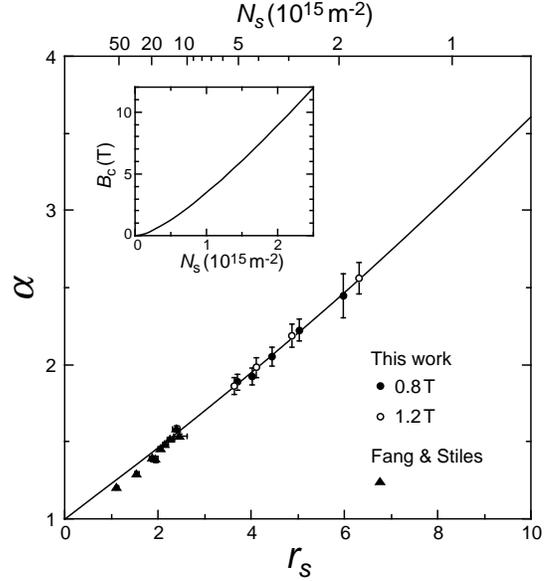}
}
\vspace{4mm}
\caption{ The enhancement factor $ \alpha = g _ { \rm FL } m _ { \rm FL } / 0.38 m _ { e } $ vs the strength of the electron-electron interaction. 
Inset shows the critical magnetic field $ B _ { c } $ calculated using Eq.~(\protect\ref{Hex}).
}
\label{FGAL}
\end{figure}

In Fig.~\ref{FGAL}, we show the ratio of $ g _ { \rm FL } m _ { \rm FL } $ to $ g ^ { \ast } m ^ { \ast } $ in the non-interacting system.
The enhancement factor $ \alpha = g _ { \rm FL } m _ { \rm FL } / 0.38 m_ { e } $ increases monotonically with the strength of the {\it e-e} interaction.
Results on much lower mobility samples measured in higher magnetic fields by Fang and Stiles \cite{FS} are also plotted.
No $ B _ { \perp } $-dependence is found between our results at $ 0.8~{ \rm T } $ and at $ 1.2~{ \rm T } $.
We believe that these values of $ B _ { \perp } $ are small enough and $ g _ { \rm FL } m _ { \rm FL } $ was determined in the limit of $ B _ { \perp } = 0 $.
While $ B _ { \perp } $ may cause a small oscillation of the effective $ g $-factor with the maxima $ g _ { \rm FL } ^ { \rm MAX } $ at $ \nu / 2 = odd $ and the minima $ g _ { \rm FL } ^ { \rm MIN } $ at $ \nu / 2 = even $ \cite{KOBAYASHI}, $ g _ { \rm FL } m _ { \rm FL } $ determined from the critical angle for $ \Delta E _ { odd } = \Delta E _ { even } $ is related to the average of $ g _ { \rm FL } ^ { \rm MAX } $ and $ g _ { \rm FL } ^ { \rm MIN } $.
The present method cannot be used in the large-$ r _ { s } $ region where $ g _ { \rm FL } m _ { \rm FL } / m _ { e } $ exceeds unity ($ \alpha > 2.63 $).
We use a simple extrapolation function 
\begin{eqnarray}
\alpha - 1 = 0.2212 r _ { s } + 0.003973 r _ { s } {} ^ { 2 } .
\label{Hex}
\end{eqnarray}
represented by the solid line in Fig.~\ref{FGAL}.
The critical magnetic field $ B _ { c } = 2 \pi \hbar ^ { 2 } N _ { s } / \mu _ { B } g _ { v } g _ { \rm FL } m _ { \rm FL } $ is calculated using (\ref{Hex}) as shown in the inset to Fig.~\ref{FGAL}.

\begin{figure}
\begin{center}
\epsfysize=71mm
\epsfbox{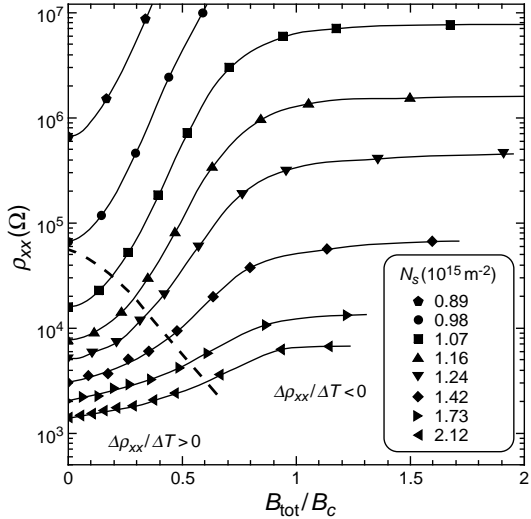}
\end{center}
\vspace{4mm}
\caption{ Resistivity vs $ B _ { \rm tot } / B _ { c } $ at $ T = 0.21~{ \rm K } $ and $ B _ { \perp } = 0 $ for various $ N _ { s } $.
Dashed line represents the critical value of $ \rho _ { xx } $ at the MIT determined from temperature dependence between 0.21~K and 0.91~K.}
\label{FGPARA}
\end{figure}

In Fig.~\ref{FGPARA}, $ \rho _ { xx } $ at $ T = 0.21~{ \rm K } $ in the parallel magnetic field ($ B _ {\perp} = 0 $) is shown.
The diagonal resistivity $ \rho _ { xx } $ increases with $ B _ { \rm tot } $ in the low-$ B _ { \rm tot } $ region, but it takes almost constant values in the high-$ B _ { \rm tot } $ region.
The saturation of $ \rho _ { xx } $ occurs around $ B _ { \rm tot } = B _ { c } $ at which the spin polarization is expected to be completed.
The result indicates that the mixing of the different spin states causes the reduction of $ \rho _ { xx } $ in the low-$ B _ { \rm tot } $ region.
The dashed line represents the critical value of $ \rho _ { xx } $ at the MIT.
It was tentatively determined from the sign of the change in $ \rho _ { xx } $ from $ T = 0.21~{ \rm K } $ to $ 0.91~{ \rm K } $.
The positive temperature dependence of $ \rho _ { xx } $ below $ 1~{ \rm K} $ was not observed for $ B _ { \rm tot } > B _ { c } $.
The positive $ T $-dependence of $ \rho _ { xx } $ in the low-$ B _ { \rm tot } $ region may arise from the scattering related to the spin degree of freedom.

\begin{figure}
\centerline{
\epsfysize=75mm
\epsfbox{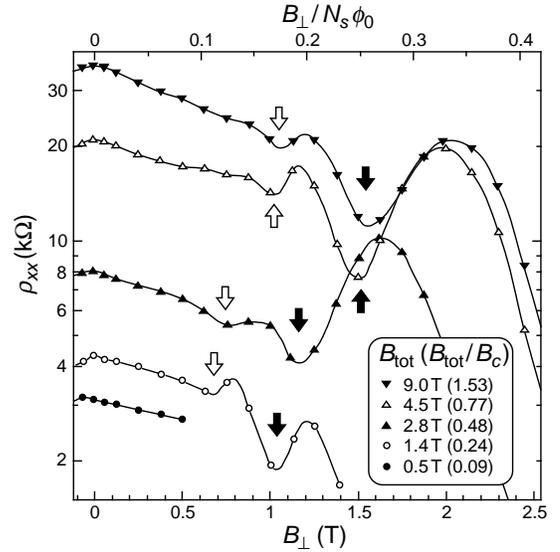}
}
\vspace{4mm}
\caption{ Diagonal resistivity vs normal component of the magnetic field at $ T = 0.31~{\rm K} $ and $ N _ { s } = 1.46 \times 10 ^ { 15 } { \rm m } ^ { -2 } $ for several values of $ B _ { \rm tot } $.}
\label{FGROT}
\end{figure}

The most important finding has been obtained by rotating the sample in the magnetic field.
Results at $ N _ { s } = 1.46 \times 10 ^ { 15 } {\rm m } ^ { -2 } $ are shown in Fig.~\ref{FGROT}.
We focus on the $ \rho _ { xx } $ oscillation that appears in the low-$ B _ { \perp } $ region, while the $ \rho _ { xx } ( B _ { \perp } ) $ curves for different $ B _ { \rm tot } $ approach each other in the higher-$ B _ { \perp } $ region (not shown in Fig.~\ref{FGROT}) and converge on a single curve for $ B _ { \perp } / N _ { s } \phi _ { 0 } > 0.5 $ with two deep minima at $ B _ { \perp } / N _ { s } \phi _ { 0 } = 0.5 $ ($ \nu = 2 $) and $ B _ { \perp } / N _ { s } \phi _ { 0 } = 1.0 $ ($ \nu = 1 $) as reported by Kravchenko {\it et al.} \cite{TILT}.
Unlike ordinary Shubnikov-de Haas oscillations that show $ \rho _ { xx } $ minima at fixed points with $ \nu = { \rm integer } $, the $ \rho _ { xx } $ minima indicated by the black arrows and the white arrows in Fig.~\ref{FGROT} shift toward high-$ B _ { \perp } $ side as $ B _ { \rm tot } $ increases.
Such peculiar behavior was also observed for lower-$ N _ { s } $ down to $ 1.02 \times 10 ^ { 15 } {\rm m } ^ { -2 } $.

\begin{figure}
\centerline{
\epsfysize=73mm
\epsfbox{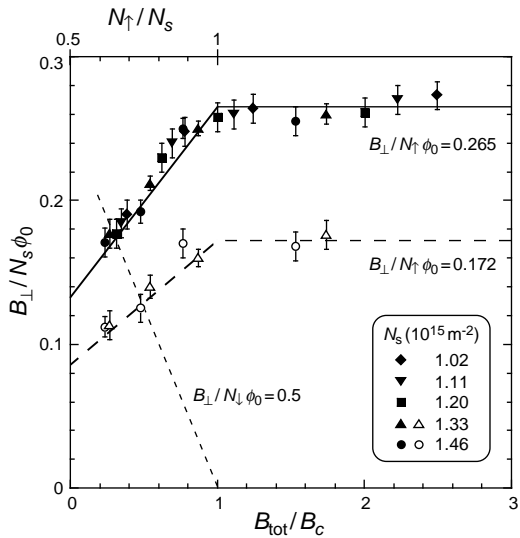}
}
\vspace{4mm}
\caption{ Positions of the $ \rho _ { xx } $ minima for various $ N _ { s } $ are plotted.
Solid line and dashed line represent $ B _ { \perp } / N _ { \uparrow } \phi _ { 0 } = 0.265 $ and $ B _ { \perp } / N _ { \uparrow } \phi _ { 0 } = 0.172 $, respectively.
Dotted line represents $ B _ { \perp } / N _ { \downarrow } \phi _ { 0 } = 0.5 $.
}
\label{FGMIN}
\end{figure}

In Fig.~\ref{FGMIN}, the positions of the $ \rho _ { xx } $ minima for various $ N _ { s } $ are shown.
Both $ B _ { \perp } $ and $ B _ { \rm tot } $ are normalized by $ N _ { s } \phi _ { 0 } $ and $ B _ { c } $, respectively.
The closed (open) symbols correspond to the black (white) arrows in Fig.~\ref{FGROT}.
The value of $ B _ { \perp } / N _ { s } \phi _ { 0 } $ increases linearly with $ B _ { \rm tot } / B _ { c } $ for $ B _ { \rm tot } / B _ { c } < 1 $, but saturates for $ B _ { \rm tot } / B _ { c } > 1 $ where the spin polarization is expected to be completed.
We link this behavior to the concentration $ N _ { \uparrow } $ of electrons having an up spin.
Assuming that a change in $ p $ by $ B _ { \perp } $ is negligible in the low-$ B _ { \perp } $ region, we have $ N _ { \uparrow } / N _ { \rm s } = (1 + B _ { \rm tot } / B _ { c } ) / 2 $ for $ B _ { \rm tot } / B _ { c } < 1 $ and $ N _ { \uparrow } / N _ { \rm s } = 1 $ for $ B _ { \rm tot } / B _ { c } > 1 $.
All the data are distributed along the solid line or the dashed line on which the number of the flux quantum per ``spin-up'' electron takes a constant value of $ B _ { \perp } / N _ { \uparrow } \phi _ { 0 } = 0.265 $ or $ 0.172 $.

One may attribute the observed $ \rho _ { xx } $ minima to the Shubnikov-de Haas oscillation produced by spin-up electrons.
Actually, the effective Landau level filling factor of spin-up electrons $ N _ { \uparrow } \phi _ { 0 } / B _ { \perp } $ at the $ \rho _ { xx } $ minima is close to an even number of 4 or 6.
Note that we have the valley degeneracy $ g _ { v } = 2 $ in this system.
It is expected that $ \rho _ { xx } $ also takes minima at even-numbers of the effective Landau level filling factor of spin-down electrons.
However, we observed no feature even for $ N _ { \downarrow } \phi _ { 0 } / B _ { \perp } = 2 $ represented by the dotted line in Fig.~\ref{FGMIN}.
The disappearance of $ \rho _ { xx } $ minima related to spin-down electrons implies that spin-down electrons do not contribute to the conduction owing to localization or large scattering rate.
If that is the case, one should accept an unusual negative dependence of the mobility of spin-up electrons on the carrier concentration $ N _ { \uparrow } $ since the value of $ \rho _ { xx } $ steeply increases with the spin polarization.
The $ \rho _ { xx } $ minimum at $ B _ { \perp } / N _ { \uparrow } \phi _ { 0 } \approx 0.265 $ appears even in the insulating region with $ \rho _ { xx } \sim 10 ^ { 6 } ~ \Omega $, where the Landau level separation is considered to be entirely smeared out by the level broadening.
Further consideration is required to explain the results consistently.

Finally, we also propose a quite different interpretation of the $ \rho _ { xx } $ minima.
In the previous work \cite{OKA1,OKA2}, we have investigated the activation energy in the insulating phase which depends on $ B _ { \perp } $ and $ B _ { \rm tot } $.
Oscillatory behavior linked with $ B _ { \perp } / N _ { s } \phi _ { 0 } $ was explained using a model based on the Aharonov-Bohm effect on the exchange of electrons localized in the pinned Wigner solid \cite{RVD}.
This model may be partly applied to the present case if one assumes that spin-up electrons are localized and form an ordered state in the real space like a Wigner solid.
Then, the $ B _ { \rm tot } $-dependence of $ \rho _ { xx } $ can be understood as a result of the change in the concentration of spin-down electrons which carry the charge.
While we at this stage do not identify the origin of the $ \rho _ { xx } $ minima specifically, we consider that the effect of the magnetic flux on the system of spin-up electrons can cause an oscillatory perturbation in the mobility of spin-down electrons through some interactions.

In summary, we have studied the magnetotransport in a high mobility Si-MOSFET at electron concentrations around the MIT.
The relationship between the resistivity and the spin polarization indicates that the mixing of the different spin states is important for the metallic conduction.
We also found the $ \rho _ { xx } $ minima in the $ B _ { \perp } $-dependence, which should be related to the concentration of ``spin-up'' electrons.

This work is supported in part by Grants-in-Aid for Scientific Research from the Ministry of Education, Science, Sports and Culture, Japan, and High-Tech-Research Center in Gakushuin University.


\vspace{-3mm}

\begin{references}

\vspace{-10mm}

\bibitem{KRAV1}
S. V. Kravchenko, G. V. Kravchenko, J. E. Furneaux, V. M. Pudalov, and M. D'Iorio, Phys. Rev. B {\bf 50}, 8039 (1994); S. V. Kravchenko, W. E. Mason, G. E. Bowker, J. E. Furneaux, V. M. Pudalov, and M. D'Iorio, Phys. Rev. B {\bf 51}, 7038 (1995); S. V. Kravchenko, D. Simonian, M. P. Sarachik, W. Mason, and J. E. Furneaux, Phys. Rev. Lett. {\bf 77}, 4938 (1996); D. Simonian, S. V. Kravchenko, and M. P. Sarachik, Phys. Rev. B {\bf 55}, 13421 (1997).

\bibitem{POPO}
D. Popovi\'{c}, A. B. Fowler, and S. Washburn, Phys. Rev. Lett. {\bf 79}, 1543 (1997).

\bibitem{SIGE1}
J. Lam, M. D'Iorio, D. Brown, and H. Lafontaine, Phys. Rev. B {\bf 56}, 12741 (1997).

\bibitem{SIGE2}
P. T. Coleridge, R. L. Williams, Y. Feng, and P. Zawadzki, Phys. Rev. B {\bf 56} 12764 (1997).

\bibitem{GAAS1}
Y. Hanein, U. Meirav, D. Shahar, C. C. Li, D. C. Tsui, and H. Shtrikman, Phys. Rev. Lett. {\bf 80}, 1288 (1998).

\bibitem{GAAS2}
M. Y. Simmons, A. R. Hamilton, M. Pepper, E. H. Linfield, P. D. Rose, and D. A. Ritchie, Phys. Rev. Lett. {\bf 80}, 1292 (1998).

\bibitem{GANG4}
E. Abrahams, P. W. Anderson, D. C. Licciardello, and T. V. Ramakrishnan, Phys. Rev. Lett. {\bf 42}, 673 (1979).

\bibitem{KRAV2}
D. Simonian, S. V. Kravchenko, M. P. Sarachik, and V. M. Pudalov, Phys. Rev. Lett. {\bf 79}, 2304 (1997); Phys. Rev. B {\bf 57}, 9420 (1998).

\bibitem{PUD1}
V. M. Pudalov, G. Brunthaler, A. Prinz, and G. Bauer, Pis'ma Zh. \'{E}ksp. Teor. Fiz. {\bf 65}, 887 (1997) [JETP Lett. {\bf 65}, 932 (1997)].

\bibitem{PUD2}
V. M. Pudalov, M. D'Iorio, S. V. Kravchenko, and J. W. Campbell, Phys. Rev. Lett. {\bf 70}, 1866 (1993).

\bibitem{CHUI}
S. T. Chui and B. Tanatar, Phys. Rev. Lett. {\bf 74}, 458 (1995); Phys. Rev. B {\bf 55}, 9330 (1997).

\bibitem{OKA1}
T. Okamoto and S. Kawaji, J. Phys. Soc. Jpn. {\bf 65}, 3716 (1996).

\bibitem{OKA2}
T. Okamoto and S. Kawaji, Phys. Rev. B {\bf 57}, 9097 (1998).

\bibitem{HT}
At higher temperatures above $ T \sim 2~{\rm K} $, positive temperature dependence appears at $ 9~{\rm T} $ in the low-$ \rho _ { xx } $ region.

\bibitem{KODERA}
H. Kodera, J. Phys. Soc. Jpn. Suppl. {\bf 21}, 578 (1966).

\bibitem{FS}
F. F. Fang and P. J. Stiles, Phys. Rev. {\bf 174}, 823 (1968).

\bibitem{KOBAYASHI}
M. Kobayashi and K. F. Komatsubara, Jpn. J. Appl. Phys. Suppl. 2, Pt. 2, 343.

\bibitem{TILT}
S. V. Kravchenko, D. Simonian, M. P. Sarachik, A. D. Kent, and V. M. Pudalov, Phys. Rev. B { \bf 58}, 3553 (1998).

\bibitem{RVD}
The model used in Ref.~\protect\onlinecite{OKA2} should be partly modified because the value of the bare valley splitting energy was overestimated to be $ \Delta E _ { v } \approx 4 ~ { \rm K } $.
Since no indication of the valley splitting was observed in the Shubnikov-de Haas oscillation as shown in Fig.~\protect\ref{FGOS}, we at present evaluate the bare valley splitting energy to be below 0.5~K for $ N _ { s } \leq 5 \times 10 ^ { 15 } { \rm m } ^ { -2 } $.
On the other hand, the theoretical values of $ \Delta E _ { v } $ have large uncertainties as reviewed by T. Ando, A. B. Fowler, and F. Stern in Rev. Mod. Phys. {\bf 54}, 437 (1982).

\end{references}
\end{document}